# Unusual hydrogen atom display in solid acids

**Boris V. Merinov**

*California Institute of Technology, Materials and Process Simulation Center, MC 139-74, Pasadena, CA 91125, USA. Correspondence e-mail: merinov@wag.caltech.edu*

Studying crystal structures of superprotonic phases of alkali metal hydrogen sulfates and selenates, a very unusual phenomenon has been revealed. Dynamically disordered hydrogen atoms with low position occupancies are clearly seen in corresponding electron density maps. To explain this effect, an idea about a new type of twinning, *dynamic twinning*, was proposed and discussed.

## 1. Introduction

Any crystallographer knows that due to the lowest scattering factor localization of hydrogen atoms using X-ray diffraction data is a non-trivial task, especially if the hydrogen atoms are disordered. Normally, it would be considered very difficult to identify electron density related to disordered hydrogen atoms, particularly, in the presence of heavy atoms such as Cs or Rb. Nevertheless, studying crystal structures of superprotonic phases of so-called solid acids (alkali metal hydrogen sulfates, selenates and phosphates), for the first time a very unusual phenomenon, which we call "*the effect of the anomalous display of dynamically disordered hydrogen atoms in electron-density maps*", has been observed (Merinov, 1997). Due to this effect, the dynamically disordered hydrogen atoms with very low position occupancies are clearly seen in electron density maps. Our detailed analysis of the electron-density distributions in the superprotonic phases of $MHAO_4$, $M_3H(AO_4)_2$ and $M_5H_3(AO_4)_4 \cdot xH_2O$, where $M=NH_4$, K, Rb, Cs and A=S, Se, revealed the presence of electron density peaks which correspond to dynamically disordered hydrogen atoms with position occupancies of 1/6, 1/12 and even less (Merinov, 1997; Merinov & Bismayer, 2000). Moreover, coordinates and thermal parameters of these disordered hydrogen atoms can be refined (Lukaszewicz *et al.*, 1993; Merinov, 1997; Merinov *et al.*, 2000; Fukami *et al.*, 2000; Merinov *et al.*, 2002). It should be noted that quite often the thermal parameters of the disordered hydrogen atoms have surprisingly low positive or even negative values. The obtained results show that the unusual or, as it was called in previous papers, "anomalous" display of the hydrogen atoms in the corresponding electron density maps appears to be a characteristic feature of the superprotonic phases of solid acid proton conductors. Indeed, this effect has already been observed for $CsHSO_4$ (Merinov, 1997), $Cs_3H(SeO_4)_2$ (Merinov, 1997; Merinov & Bismayer, 2000), $(NH_4)_3H(SeO_4)_2$ (Lukaszewicz *et al.*, 1993), $(ND_4)_3D(SO_4)_2$ (Fukami *et al.*, 2000), $Rb_3H(SeO_4)_2$ (Merinov *et al.*, 2002), $[Rb_{0.57}(NH_4)_{0.43}]_3H(SeO_4)_2$ (Merinov *et al.*, 2000), and $Cs_5H_3(SO_4)_4 \cdot xH_2O$ (Merinov, 1997). It would seem such an extraordinary phenomenon should attract enhanced attention of researchers. However, up to the present the only attempt, evolving an idea about a new type of twinning, which we call *dynamic twinning* (Merinov *et al.*, 2000), was undertaken to explain this unusual effect observed in the superprotonic phases of solid acids.

## 2. Dynamic twinning as a possible reason for the unusual display of the hydrogen atoms

The superprotonic phase transitions in $MHAO_4$, $M_3H(AO_4)_2$ and $M_5H_3(AO_4)_4 \cdot xH_2O$ are simultaneously ferroelastic. On cooling from the high-temperature paraelastic phases to low-temperature ferroelastic phases, the high-symmetry axis disappears below the superprotonic phase transition temperature and ferroelastic low-symmetry domains related by the high-symmetry axis are formed. The hydrogen bond networks, which are dynamically disordered in the superprotonic phases, become ordered in the ferroelastic phases. Based on our X-ray investigations we made a supposition that most probably a highly correlated dynamic disorder of the hydrogen bonds occurred in the superprotonic phases of the solid acids (Merinov, 1996; Merinov & Bismayer, 2000; Merinov *et al.*, 2000). This means that changing the orientation of one hydrogen bond induces correlated orientation changes of other hydrogen bonds and related $AO_4$ tetrahedra within some crystal volume. The size of this volume is at least of an X-ray beam coherence length (~1 μm). The reorientations of the $HAO_4$ groups occur relatively rarely, if they are examined in the time scale of X-ray scattering (Blinc *et al.*, 1984; Dmitriev *et al.*, 1986), i.e. the residence time of the orientation states is much larger than the characteristic time of the X-ray scattering process. Therefore, the observed diffraction patterns can formally be described as resulting from twinned crystals which consist of domains related by the corresponding high-symmetry axis. This looks as if the ferroelastic domains continue existing in the paraelastic phase but in a dynamic state, i.e. they switch their orientation states in time. For instance, the paraelastic superprotonic phase of $M_3H(AO_4)_2$ crystals is usually considered being trigonal with space group $R\bar{3}m$, whereas the ferroelastic phases are monoclinic (space group $C2/m$ or $C2/c$) and three orientation states related by the three-fold symmetry axis are formed in the ferroelastic phase. From the observed trigonal symmetry and dynamic nature of the specific twinning in the high-temperature paraelastic phase follows equality of the twin component volumes. In this particular case, the difference between the hydrogen contributions to the reflection intensities of the twinned monoclinic and untwinned trigonal structures can be expressed as follows

$$\Delta = |f^H \exp(i2\pi \mathbf{H}\mathbf{r}_j)|^2 - |\sum_{j=1}^{3} 1/3 f^H \exp(i2\pi \mathbf{H}\mathbf{r}_j)|^2, \quad (1)$$

where $f^H$ is the atomic scattering factor of hydrogen and $\mathbf{r}_j$ is a three-dimensional position vector of the hydrogen atom. This may result in "amplification" of the electron density at the acid hydrogen atom position in the untwinned trigonal model (Merinov et al., 2000).

### 3. Unusual display of hydrogen atoms

Figure 1 shows an example of the unusual display of the dynamically disordered hydrogen atoms in the superprotonic phase of $Cs_3H(SeO_4)_2$ (space group $R\bar{3}m$). Three acid hydrogen atoms occupy the 18(h) crystallographic position with a probability of 1/6. This dynamic disorder includes 1) disordering of the hydrogen atom within a double-well potential of the hydrogen bond and 2) disordering of the overall hydrogen bond network (only 1/3 of possible hydrogen bond positions are occupied). In Figure 1 one can clearly see details of the hydrogen transfer within the two-minimum hydrogen bond in $Cs_3H(SeO_4)_2$. During the $SeO_4$-group librations the hydrogen bond shortens and elongates as though it breathes. In accordance with this, the hydrogen atom moves within the double-well potential of the corresponding symmetric hydrogen bond. For instance, let's assume that we have the O–H⋯O hydrogen bond. As mentioned above, during the $SeO_4$-group librations the H-atom could shift towards the center of the hydrogen bond and when the bond is short enough and, therefore, the barrier between two minima of the hydrogen bond is low, the H-atom could move to the other minimum (this hydrogen transfer is shown by an arrow in Figure 1) and switch the donor to the acceptor and vice versa to form the O⋯H–O hydrogen bond. If we proceed from the assumption that the dynamic twinning takes place in the $Cs_3H(SeO_4)_2$ superprotonic phase, then in each monoclinic orientation state only the disordering of the hydrogen within the two-minimum hydrogen bond remains and the hydrogen position occupancy becomes equal to 1/2.

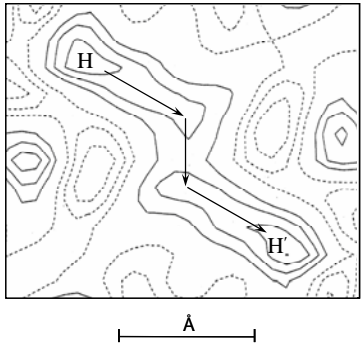

**Figure 1.** High-temperature (470 K) superprotonic phase of $Cs_3H(SeO_4)_2$ (Merinov & Bismayer, 2000). Difference electron density map after refinement of all non-hydrogen atoms. Contour intervals are drawn at steps of 0.1 $e\text{Å}^{-3}$, positive contours are represented by solid lines, negative contours by dashed lines and zero levels are omitted.

Figure 2 shows another example of the unusual display of the dynamically disordered hydrogen atoms in the superprotonic phase of the mixed $[Rb_{0.57}(NH_4)_{0.43}]_3H(SeO_4)_2$ compound (space group $R\bar{3}m$). Despite extremely low occupancies of the hydrogen atom positions [1/12 for the acid hydrogen at the 36(i) position, and ~1/20 and 1/6 for the $NH_4$-tetrahedron top and base hydrogen atoms at the 18(h) and 18(f) positions, respectively], they are amazingly clearly seen in the corresponding electron density map. Here we can see that not only the acid hydrogen atoms, but also the ammonium hydrogen atoms are disordered. And once again, the unusual display of the hydrogen atoms allows details of the behavior of the hydrogen atoms and structural units in $[Rb_{0.57}(NH_4)_{0.43}]_3H(SeO_4)_2$ to be elicited from the corresponding electron density maps. For instance, analyzing the electron density of the ammonium hydrogen atoms, we can conclude that the $NH_4$ groups undergo a one-dimensional hindered rotation around the z-axis in the superprotonic phase of $[Rb_{0.57}(NH_4)_{0.43}]_3H(SeO_4)_2$. Similar results were obtained in spectroscopic studies of $(NH_4)_3H(SeO_4)_2$ (Kamoun et al., 1987; Tritt-Goc et al., 1993). The disorder of the acid hydrogen atoms results from the librations of the $SeO_4$-tetrahedra, to the top oxygen atoms of which these hydrogen atoms are bonded, and from the proton transfer within the two-minimum hydrogen bonds. Both of these processes manifest themselves in the electron density maps (Figure 2).

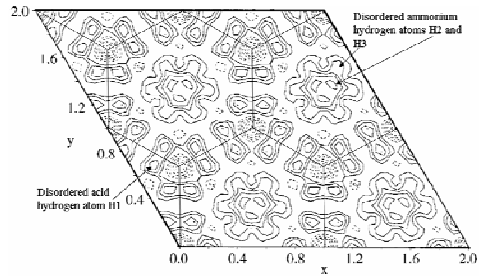

**Figure 2.** High-temperature (370 K) superprotonic phase of $[Rb_{0.57}(NH_4)_{0.43}]_3H(SeO_4)_2$ (Merinov et al., 2000). Difference electron density map in the plane of the dynamically disordered acid hydrogen bonds that form a planar hexagonal network shown by thin solid lines. Contour intervals are drawn at steps of 0.05 $e\text{Å}^{-3}$.

If we treat the $[Rb_{0.57}(NH_4)_{0.43}]_3H(SeO_4)_2$ superprotonic phase with the model based on the dynamic twinning, then the occupancies of the related disordered atomic positions increase in accordance with the number of orientation states, *i.e.* three times, and thermal parameters of the hydrogen atoms, which were negative or very small in the high-symmetry disordered structure, accept reasonable positive values (Merinov *et al.*, 2000).

It should be noted that all the aforesaid applies to the disordered atoms, which allow alternative description via dynamic twinning. This effect should not be observed and it is not observed for the atoms which completely occupy their crystallographic positions.

## 4. Concluding remarks

The main goal of this paper is to attract attention of researchers, in particular crystallographers, to the extraordinary phenomenon which is observed in superprotonic phases of alkali metal hydrogen sulfates and selenates and which we call the unusual (or anomalous) display of dynamically disordered hydrogen atoms. The essence of this phenomenon is that the dynamically disordered hydrogen atoms with very low position occupancies are clearly seen in the electron density maps. To explain this effect we have introduced a concept about dynamic twinning in the superprotonic phases of solid acids. However, dynamic twinning is rather an exciting hypothesis than a proved fact. The principle question "What causes the unusual display of the dynamically disordered hydrogen atoms?" remains open and need further investigations. Single crystals of solid acids can be relatively easily grown by slow evaporation from aqueous solutions. Any crystallographic or materials science group can collect X-ray diffraction data and reproduce the reported results.